\documentclass[epj]{svjour}
\usepackage{latexsym}
\usepackage{graphics}
\usepackage{dcolumn}

\begin{document}
\title{\emph{(n,$\gamma$)} Cross Sections of Light \emph{p} Nuclei} \subtitle{Towards an Updated Database for the p Process}
\author{I. Dillmann\inst{1}$^,$\inst{2} \and M. Heil\inst{1} \and F. K\"appeler\inst{1} \and R. Plag\inst{1}
\and T. Rauscher\inst{2} \and F.-K. Thielemann\inst{2}
}                     
\offprints{iris.dillmann@ik.fzk.de}          
\institute{Institut f\"ur Kernphysik, Forschungszentrum Karlsruhe,
Postfach 3640, D-76021 Karlsruhe, Germany \and Departement Physik
und Astronomie, Universit\"at Basel, Klingelbergstrasse 82,
CH-4056 Basel, Switzerland}
\date{Received: date / Revised version: date}
%
\abstract{The nucleosynthesis of elements beyond iron is dominated
by the \emph{s} and \emph{r} processes. However, a small amount of
stable isotopes on the proton-rich side cannot be made by neutron
capture and are thought to be produced by photodisintegration
reactions on existing seed nuclei in the so-called "\emph{p}
process". So far most of the \emph{p}-process reactions are not
yet accessible by experimental techniques and have to be inferred
from statistical Hauser-Feshbach model calculations. The
parametrization of these models has to be constrained by
measurements on stable proton-rich nuclei. A series of
(\emph{n},$\gamma$) activation measurements on \emph{p} nuclei,
related by detailed balance to the respective
photodisintegrations, were carried out at the Karlsruhe Van de
Graaff accelerator using the $^7$Li(\emph{p,n})$^7$Be source for
simulating a Maxwellian neutron distribution of \emph{kT}= 25 keV.
We present here preliminary results of our extended measuring
program in the mass range between \emph{A}=74 and \emph{A}=132,
including first experimental (\emph{n},$\gamma$) cross sections of
$^{74}$Se, $^{84}$Sr, $^{120}$Te and $^{132}$Ba, and an improved
value for $^{130}$Ba. In all cases we find perfect agreement with
the recommended MACS predictions from the Bao et al. compilation.
} 
\maketitle
\section{Introduction}
\label{intro} Astrophysical models can explain the origin of most
nuclei beyond the iron group in a combination of processes
involving neutron captures on long (\emph{s} process) or short
(\emph{r} process) time scales \cite{bbfh57,lawi01}. However, 32
stable, proton-rich isotopes between $^{74}$Se and $^{196}$Hg
cannot be formed in that way. Those \emph{p} nuclei are 10 to 100
times less abundant than the \emph{s} and \emph{r} nuclei in the
same mass region. They are thought to be produced in the so-called
$\gamma$ or \emph{p} process, where proton-rich nuclei are made by
sequences of photodisintegrations and $\beta^+$ decays
\cite{woho78,woho90,raar95}. In this scenario, pre-existing seed
nuclei from the \emph{s} and \emph{r} processes are destroyed by
photodisintegration in a high-temperature environment, and
proton-rich isotopes are produced by ($\gamma$,n) reactions. When
($\gamma$,\emph{p}) and ($\gamma,\alpha$) reactions become
comparable or faster than neutron emission within an isotopic
chain, the reaction path branches out and feeds nuclei with lower
charge number \emph{Z}. The decrease in temperature at later
stages of the \emph{p} process leads to a freeze-out via neutron
captures and mainly $\beta^+$ decays, resulting in the typical
\emph{p}-process abundance pattern with maxima at $^{92}$Mo
(\emph{N}=50) and $^{144}$Sm (\emph{N}=82).

The currently most favored astrophysical site for the \emph{p}
process is explosive burning in type II supernovae. The explosive
shock front heats the outer O/Ne shell of the progenitor star to
temperatures of 2-3 GK, sufficient for providing the required
photodisintegrations. Following the nucleosynthesis in such
astrophysical models, good agreement with the required \emph{p}
production is found, with exception of the low (\emph{A}$<$100)
and intermediate (150$\leq$\emph{A}$\leq$ 165) mass range, which
are underproduced by factors of \mbox{3-4 \cite{rahe02}}.
Currently, however, it is not yet clear whether the observed
underproductions are due to a problem with astrophysical models or
with the nuclear physics input, i.e. reaction rates. Thus, a
necessary requirement towards a consistent understanding of the
\emph{p} process is the reduction of uncertainties in nuclear
data. By far most of the several hundreds of required
photodisintegration rates and their inverses need to be inferred
from Hauser-Feshbach statistical model calculations
\cite{hafe52,rau95}. Experimental data can improve the situation
in two ways, either by directly replacing predictions with
measured cross sections in the relevant energy range, or by
testing the reliability of predictions at other energies when the
relevant energy range is not experimentally accessible.

The role of (\emph{n},$\gamma$) reactions in the \emph{p} process
was underestimated for a long time, although it is obvious that
they have an influence on the final \emph{p}-process abundances.
Neutron captures compete with ($\gamma$,\emph{n}) reactions and
thus hinder the photodisintegration flux towards light nuclei,
especially at lower-\emph{Z} isotopes and even-even isotopes in
the vicinity of branching-points. Rayet et al. \cite{ray90} have
studied the influence of several components in their
\emph{p}-process network calculations. Their nuclear flow schemes
show that branching points occur even at light \emph{p} nuclei,
and are shifted deeper into the proton-rich unstable region with
increasing mass and temperature. In contradiction to Woosley and
Howard \cite{woho78}, who claimed for their network calculations
that (\emph{n},$\gamma$) can be neglected except for the lightest
nuclei (\emph{A}$\leq$90), Rayet et al. also examined the
influence of neutron reactions for temperatures between
\emph{T}$_9$= 2.2 and 3.2 GK by comparing overabundance factors if
(\emph{n},$\gamma$) reactions on \emph{Z}$>$26 nuclides are
considered or completely suppressed. As a result, the
overabundances were found to change by up to a factor 100 (for
$^{84}$Sr) if the (\emph{n},$\gamma$) channel was artificially
suppressed. This rather high sensitivity indicates the need for
reliable (\emph{n},$\gamma$) rates to be used in \emph{p}-process
networks.

The influence of a variation of reaction rates on the final
\emph{p} abundances has also been studied previously
\cite{rau05,rapp04}. It turned out that the \emph{p} abundances
are very sensitive to changes of the neutron-induced rates in the
entire mass range, whereas the proton-induced and $\alpha$-induced
reaction rates are important at low and high mass numbers,
respectively.

A third reason for the determination of neutron capture rates of
\emph{p} nuclei are those cases where experimental
photodissociation rates are not accessible. The respective
astrophysical photodisintegration rate can then be inferred from
capture rates by detailed balance \cite{rau00}. This is the case
for most stable \emph{p} nuclei, which are separated from stable
isotopes by a radioactive nucleus. While the reaction rate
$^A$X($\gamma$,\emph{n})$^{A-1}$X can be determined by
brems\-strahlung \cite{vomo01}, the reaction
$^{A+1}$X($\gamma$,\emph{n})$^{A}$X has to be measured via its
inverse (\emph{n},$\gamma$) rate.

The present work comprises the first measurement of
(\emph{n},$\gamma$) cross sections for the \emph{p}-process
isotopes $^{74}$Se, $^{84}$Sr, $^{120}$Te and $^{132}$Ba at
\emph{kT}= 25 keV, and a re-measurement of $^{130}$Ba. The direct
determination of stellar (\emph{n},$\gamma$) rates requires a
"stellar" neutron source yielding neutrons with a
Maxwell-Boltzmann energy distribution. We achieve this by making
use of the $^7$Li(\emph{p,n})$^7$Be reaction. In combination with
the activation or time-of-flight technique, this offers a unique
tool for comprehensive studies of (\emph{n},$\gamma$) rates and
cross sections for astrophysics.

\section{Experimental procedure}
\label{exp} All measurements were carried out at the Karlsruhe 3.7
MV Van de Graaff accelerator using the activation technique.
Neutrons were produced with the $^7$Li(\emph{p,n})$^7$Be sour\-ce
by bombarding 30 $\mu$m thick layers of metallic Li on a
water-cooled Cu backing with protons of 1912 keV, 30 keV above the
reaction threshold. The resulting quasi-stellar neutron spectrum
approximates a Maxwellian distribution for \emph{kT}= \mbox{25.0
$\pm$ 0.5 keV} \cite{raty88}. Hence, the proper stellar capture
cross section can be directly deduced from our measurement.

For all activations natural samples of the respective element were
used. The selenium and tellurium samples were prepared from metal
granules, whereas for the barium and strontium measurement thin
pellets were pressed from powders of Sr(OH)$_2$, SrF$_2$, SrCO$_3$
and BaCO$_3$. In order to verify the stoichiometry, the powder
samples were dried at 300$^\circ$C and 800$^\circ$C, respectively.
All samples were enclosed in a 15 $\mu$m thick aluminium foil and
sandwiched between 10-30 $\mu$m thick gold foils of the same
diameter. In this way the neutron flux can be determined relative
to the well-known capture cross section of $^{197}$Au
\cite{raty88}.

Throughout the irradiation the neutron flux was recor\-ded in
intervals of 1~min using a $^6$Li-glass detector for later
correction of the number of nuclei, which decayed during the
activation. The activations were carried out with the Van de
Graaff accelerator operated in DC mode with a current of
$\approx$100~$\mu$A. The mean neutron flux over the period of the
activations was $\approx$1.5$\times$10$^9$ n/s at the position of
the samples, which were placed in close geometry to the Li target.
The duration of the single activations varied between 3~h (for the
partial cross section to $^{85}$Sr$^m$, t$_{1/2}$= 67.6 m) and
130~h (for determination of the 10.52~y ground-state of
$^{133}$Ba).

\section{Data analysis} \label{Data}
The induced $\gamma$-ray activities were counted after the
irradiation in a well defined geometry using a shielded Ge
detector in a low background area. Energy and efficiency
calibrations have been carried out with a set of reference
$\gamma$-sources in the energy range between 60 keV and 2000 keV.
For the counting of the long-lived $^{133}$Ba$^g$ activity
(t$_{1/2}$= 10.52 y) two fourfold segmented Clover detectors in
close geometry were used \cite{saed}.

\begin{table*}
\caption{Decay properties of the product nuclei \cite{nndc}.
Isotopic abundances are from Ref. \cite{iupac}.} \label{decay}
\begin{tabular}{cccccc}
Reaction & Isot. abund. [\%] & Final state & t$_{1/2}$ & E$_\gamma$ [keV] & I$_\gamma$ [\%]\\
\hline
$^{74}$Se(n,$\gamma$)$^{75}$Se & 0.89 (0.04) & Ground state & 119.79 $\pm$ 0.04 d & 136.0 & 58.3 $\pm$ 0.7 \\
& & & & 264.7 & 58.9 $\pm$ 0.3 \\
\hline $^{84}$Sr(n,$\gamma$)$^{85}$Sr & 0.56 (0.01) & Ground state & 64.84 $\pm$ 0.02 d & 514.0 & 96 $\pm$ 4 \\
& & Isomer & 67.63~$\pm$~0.04 m & 151.2 (EC) & 12.9 $\pm$ 0.7 \\
& & & & 231.9 (IT) & 84.4 $\pm$ 2.2 \\
\hline $^{120}$Te(n,$\gamma$)$^{121}$Te & 0.096 (0.001) & Ground state & 19.16 $\pm$ 0.05 d & 573.1 & 80.3 $\pm$ 2.5 \\
& & Isomer & 154 $\pm$ 7 d & 212.2 (IT) & 81.4 $\pm$ 1.0 \\
\hline $^{130}$Ba(n,$\gamma$)$^{131}$Ba & 0.106 (0.001) & Ground state & 11.50 $\pm$ 0.06 d & 123.8 & 29.0 $\pm$ 0.3 \\
& & & & 216.1 & 19.7 $\pm$ 0.3  \\
& & & & 373.2 & 14.0 $\pm$ 0.2 \\
& & & & 496.3 & 46.8 $\pm$ 0.2 \\
\hline $^{132}$Ba(n,$\gamma$)$^{133}$Ba & 0.101 (0.001) & Ground state & 10.52 $\pm$ 0.13 y & 356.0 & 62.1 $\pm$ 0.2 \\
& & Isomer & 38.9 $\pm$ 0.1 h & 275.9 (IT) & 17.8 $\pm$ 0.6 \\
\end{tabular}
\end{table*}

A detailed description of the analysis procedure is given in Refs.
\cite{beer80,dill05}. The number of activated nuclei \emph{A} can
be written as
\begin{eqnarray}
A(i) = \Phi_{tot}~N_i~\sigma_i~f_b(i) \label{eq:four},
\end{eqnarray}
where $\Phi_{tot} = \int \phi(t)dt$ is the time-integrated neutron
flux and \emph{N}$_i$ the number of atoms in the sample. The
factor f$_b$ accounts for the decay of activated nuclei during the
irradiation time \emph{t}$_a$ as well as for variations in the
neutron flux. As our measurements are carried out relative to
$^{197}$Au as a standard, the neutron flux $\Phi_{tot}$ cancels
out:
\begin{eqnarray}
\frac{A(i)}{A(Au)}=
\frac{\sigma_i~N_i~f_b(i)}{\sigma_{Au}~N_{Au}~f_b(Au)} \nonumber\\
\Longleftrightarrow  \sigma_i =
\frac{A(i)~\sigma_{Au}~N_{Au}~f_b(Au)}{A(Au)~N_i~f_b(i)}
\label{eq:five}.
\end{eqnarray}
The reference value for the experimental $^{197}$Au cross section
in the quasi-stellar spectrum of the $^{7}$Li(\emph{p,n})$^{7}$Be
source is 586 $\pm$ 8~mb \cite{raty88}. By averaging the induced
activities of the gold foils, one can determine the neutron flux
$\Phi$$_{tot}$ at the position of the sample and deduce the
experimental cross section $\sigma_i$ of the investigated sample
as shown in Eq.~\ref{eq:five}.

\section{Results and Discussion}
\subsection{General}
In an astrophysical environment with temperature \emph{T}, the
neutron spectrum corresponds to a Maxwell-Boltzmann distribution
\begin{eqnarray}
\Phi \sim E_n ~e^{-E_n /kT} \label{eq:six}.
\end{eqnarray}

The experimental neutron spectrum of the $^7$Li(\emph{p,n})$^7$Be
reaction approximates a Maxwellian distribution with \emph{kT}= 25
keV almost perfectly \cite{raty88}. But to obtain the exact
Maxwellian averaged cross section
$<$$\sigma$$>_{kT}$=$\frac{<\sigma\upsilon>}{\upsilon_T}$ for the
temperature \emph{T}, the energy-dependent cross section
$\sigma$(E) has to be folded with the experimental neutron
distribution to derive a normalization factor
NF=$\frac{\sigma}{\sigma_{exp}}$. The normalized cross section in
the energy range 0.01$\leq$E$_n$$\leq$4000 keV was used for
deriving the proper MACS as a function of thermal energy
\emph{kT}:
\begin{eqnarray}
\frac{<\sigma\upsilon>}{v_T}=<\sigma>_{kT}=\frac{2}{\sqrt{\pi}}
\frac{\int_{0}^{\infty} \frac{\sigma(E_n)}{NF}~E_n~e^{-E_n /kT}
~dE_n}{\int_{0}^{\infty} E_n~e^{-E_n /kT}~dE_n} \label{eq:seven}.
\end{eqnarray}
In this equation, $\frac{\sigma(E_n)}{NF}$ is the normalized
energy-dependent capture cross section and E$_n$ the neutron
energy. The factor $\upsilon_T$= $\sqrt{2kT/m}$ denotes the most
probable velocity with the reduced mass \emph{m}.

Maxwellian averaged cross sections have to be corrected by a
temperature-dependent stellar enhancement factor
\begin{eqnarray}
SEF(T) = \frac{\sigma^{star}}{\sigma^{lab}} \label{eq:eight}.
\end{eqnarray}
The stellar cross section $\sigma^{star}$=$\sum_{\mu} \sum_{\nu}
\sigma^{\mu\nu}$ accounts for all transitions from excited target
states $\mu$ to final states $\nu$ in thermally equilibrated
nuclei, whereas the laboratory cross section
$\sigma^{lab}$=$\sum_{0} \sum_{\nu} \sigma^{0\nu}$ includes only
captures from the target ground state. In the investigated cases
the thermal population effects in the stellar plasma at
\emph{p}-process temperatures are small for Se and Sr, but
increase up to 1.42 for Te and Ba (Table~\ref{sef}).
\begin{table}
\caption{Stellar enhancement factors for different temperatures
\cite{rau00}.} \label{sef}
\begin{tabular}{ccccccc}
T & \emph{kT} & SEF & SEF & SEF & SEF & SEF \\
 $[GK]$ & $[keV]$ & $^{74}$Se & $^{84}$Sr &$^{120}$Te & $^{130}$Ba & $^{132}$Ba \\
\hline
0.3 &  26 & 1.00 & 1.00 & 1.00 & 1.00 & 1.00 \\
2.0 & 172 & 1.01 & 1.02 & 1.10 & 1.23 & 1.16 \\
2.5 & 215 & 1.02 & 1.06 & 1.18 & 1.33 & 1.22 \\
3.0 & 260 & 1.03 & 1.09 & 1.25 & 1.42 & 1.28 \\
\end{tabular}
\end{table}

\subsection{Experimental results} \label{results}
For sample characteristics, activation features, and a detailed
discussion of the Se and Sr results see Ref.~\cite{dill05}. The
results of the Te and Ba measurements in this paper are yet
preliminary and correspond only to the cross sections derived with
the experimental neutron distribution at \emph{kT}= 25 keV.
Nevertheless, this value approximates the Maxwellian averaged
cross section at \emph{kT}= 30 keV very well and can be used for
comparison with other stellar cross sections. The resulting MACS
at 30 keV (for Se and Sr) and the experimental values for Te and
Ba are shown in Table~\ref{macs}. The extrapolation to higher
(\emph{p}-process) temperatures will not be discussed here and can
also be found in Ref. \cite{dill05}.

\subsubsection{$^{74}$Se(n,$\gamma$)$^{75}$Se}\label{se}
The $^{74}$Se(\emph{n},$\gamma$)$^{75}$Se reaction was analyzed
via the two stron\-gest transitions in $^{75}$As at 136.0 keV and
264.7 keV. The capture cross section derived with the experimental
neutron distribution is 281 $\pm$ 15~mb. The result for the
stellar cross section is $<$$\sigma$$>$$_{30}$= 271~mb, in perfect
agreement with the previously estimated value of 267 $\pm$ 25~mb
from Ref.~\cite{bao00}.

\subsubsection{$^{84}$Sr(n,$\gamma$)$^{85}$Sr$^{g,m}$}\label{sr}
In case of $^{84}$Sr, neutron captures populate both, ground and
isomeric state of $^{85}$Sr. While $^{85}$Sr$^{g}$ decays can be
identified via the 514 keV transition in $^{85}$Rb, the decay of
the isomer proceeds mainly via transitions of 232 keV and 151 keV.
The isomeric state is 239 keV above the ground state and decays
either via a 7 keV- 232 keV cascade (internal transition, 86.6\%)
or directly by electron capture (13.4\%) into the 151 keV level of
the daughter nucleus.

The partial cross section to the isomeric state can be easily
deduced from the above mentioned transitions at 151 keV and 232
keV and yields 189 $\pm$ 10 mb. The cross section to the ground
state was measured to 112 $\pm$ 8 mb, which leads to a total
capture cross section of 301 $\pm$ 17 mb.

The result for the total stellar cross section of $^{84}$Sr is
$<$$\sigma$$>$$_{30}$= 300 mb, 17 \% lower than the 368 $\pm$125
mb from Ref.~\cite{bao00}. The partial cross section to the isomer
yields $<$$\sigma$$>$$_{30}$(part)= 190~mb.

\subsubsection{$^{120}$Te(n,$\gamma$)$^{121}$Te$^{g,m}$}\label{te}
The Te samples were analyzed via the 576 keV $\gamma$-line from
the $\beta^+$ decay of $^{121}$Te$^g$ into $^{121}$Sb. The partial
cross section of the isomeric state cannot be measured directly
after the irradiation due to a huge Compton background around 210
keV, but after a waiting time of 80~d the expected 212 keV from
the IT to the ground state (88.6 \%) could be determined.

The preliminary result for the neutron capture to the ground-state
is 390 $\pm$ 16 mb, and 61 $\pm$ 2 mb for the partial cross
section to the isomeric state. This leads to a (preliminary) total
(\emph{n},$\gamma$) cross section of 451 $\pm$ 18 mb, which again
is in good agreement with the estimated 420 $\pm$ 103 mb from
Ref.~\cite{bao00}.

\subsubsection{$^{130}$Ba(n,$\gamma$)$^{131}$Ba}\label{ba1}
The $^{130}$Ba cross section can be determined via the transitions
at 124 keV, 216 keV, 373 keV and 496 keV from the $\beta^+$ decay
into $^{131}$Cs. The resulting experimental cross section is 694
$\pm$ 20 mb, which exhibits a much smaller uncertainty than the
760 $\pm$ 110 mb from Ref. \cite{brad79}, which were derived at a
filtered neutron beam.

\subsubsection{$^{132}$Ba(n,$\gamma$)$^{133}$Ba$^{g,m}$}\label{ba2}
The partial cross section to the 38.9~h isomer in $^{133}$Ba was
measured via the 276 keV line (IT) to be 33.6 $\pm$ 1.7~mb. The
total capture cross section was determined with a Clover detector
via the strongest EC decay transition into $^{133}$Cs at 356.0
keV. The preliminary result is 368 $\pm$ 25 mb, in perfect
agreement with the estimated 379 $\pm$ 137 mb from Ref.
\cite{bao00}.

\subsection{Comparison with theory}
\begin{figure}
\includegraphics{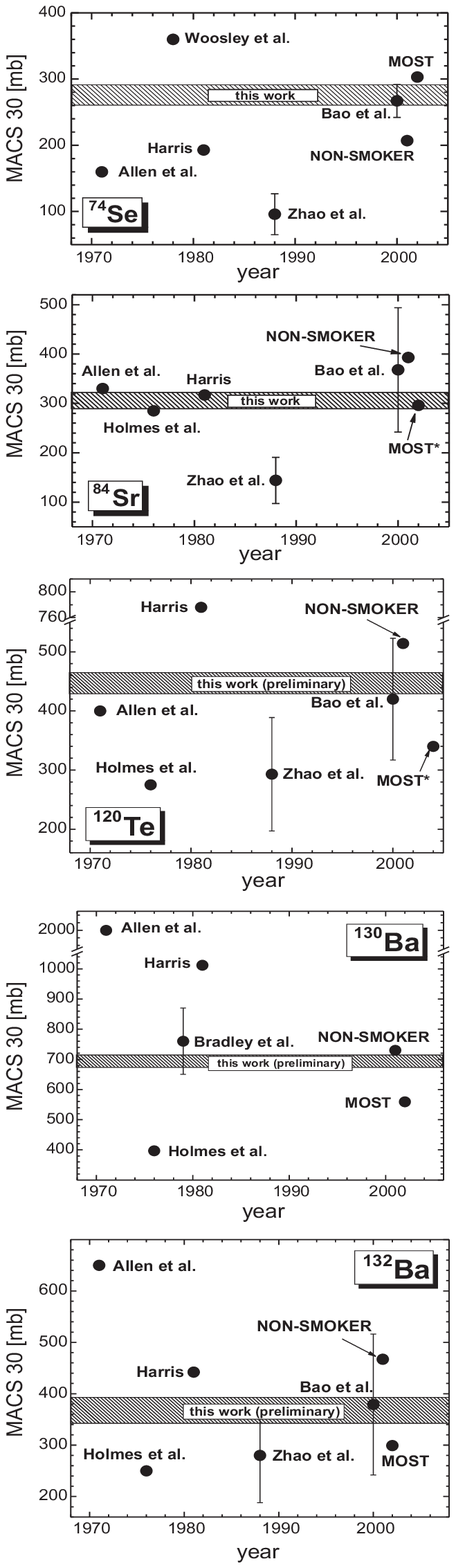}
\caption{Comparison of MACS30 predictions. MOST* marks modified
values \cite{gori05}.}
\label{macs2}       
\end{figure}

Fig.~\ref{macs2} shows a comparison of our experimental total
capture cross sections with $<$$\sigma$$>$$_{30}$ values derived
with various theoretical models
\cite{alle71,holm76,woos78,harr81,zhao88,rau01,most,gori05}. For
$^{74}$Se and $^{84}$Sr the experimental value shown is the MACS
derived with the energy dependence of JEFF 3.0 \cite{dill05,jeff},
whereas the preliminary values for $^{120}$Te, $^{130}$Ba and
$^{132}$Ba shown here are only cross sections derived with the
experimental neutron distribution.

In the case of $^{130}$Ba our experimental value agrees with the
measurement of Ref.~\cite{brad79}. In all other cases we find good
agreement with the semi-theoretical values of Bao et al.
\cite{bao00}, which are normalized NON-SMOKER cross sections
accounting for known systematic deficiencies in the nuclear inputs
of the calculation.

\section{Summary} \label{sum}
We have presented the results of an ongoing experimental program
to determine more precise \emph{p}-process reaction rates in the
mass range \emph{A}=70-140. The (\emph{n},$\gamma$) cross sections
of the \emph{p} nuclei $^{74}$Se, $^{84}$Sr, $^{120}$Te,
$^{132}$Ba have been measured for the first time, including the
partial cross sections to the isomeric states in $^{85}$Sr,
$^{121}$Te, and $^{133}$Ba. A re-measurement of $^{130}$Ba yielded
a more precise total cross section compared to the previous value
\cite{brad79}.

As can be seen in Table~\ref{macs}, experimental Maxwellian
averaged cross sections for $^{98}$Ru, $^{102}$Pd, $^{138}$La,
$^{158}$Dy, $^{168}$Yb, $^{174}$Hf, $^{184}$Os and $^{196}$Hg are
still missing. Thus, future efforts should be focussed on these
measurements, as well as on an improvement of the accuracy of
important isotopes like $^{92}$Mo and $^{94}$Mo.

\section{KADoNiS- The Karlsruhe Astrophysical Database of
Nucleosynthesis in Stars} \label{kadonis} The KADoNiS project is
an online database for cross sections in the \emph{s} process and
\emph{p} process (http://nuclear-astrophysics.fzk.de/kadonis/).
Its first part consists of an updated version of the Bao et al.
compilation \cite{bao00} for cross sections relevant to the
\emph{s} process. A test launch of the KADoNiS webpage started in
May 2005 with an online version of the original Bao et al. paper.
By end of June 2005 the first updated version was online. For the
six isotopes $^{128-130}$Xe, $^{147}$Pm, $^{151}$Sm and
$^{180}$Ta$^m$ the previously recommended semi-theoretical MACS
were replaced by first experimental results. More than 40 isotopes
(a list is available online) exhibit new measurements, which were
included to re-evaluate the recommended MACS.

The KADoNiS data sheets include all necessary information for the
respective (\emph{n},$\gamma$) reaction (recommended total and
partial cross sections, all available published values with
references, energy dependence of the MACS for 5$<$\emph{kT}$<$100
keV, and the respective stellar enhancement factors).

The second part of KADoNiS is planned to be a collection of
experimental \emph{p}-process reaction rates, including
(\emph{n},$\gamma$), (\emph{p},$\gamma$), ($\alpha$,$\gamma$) and
their respective photodissociation rates. The projected launch of
this part of the database will be December 2005.

\begin{table*}[!htb]
\caption{Maxwellian averaged cross sections $<$$\sigma$$>$$_{30}$
of all 32 \emph{p}-process nuclei at \emph{kT}= 30 keV. Values
taken from this work are in bold. ($^1$) Relative to Si $\equiv$
10$^{6}$. ($^2$) Rescaled NON-SMOKER cross sections accounting for
known systematic deficiencies in the nuclear inputs \cite{bao00}.
($^3$) Xe abundances taken from Ref.~\cite{reif02}. ($^4$)
Modified values \cite{gori05}. ($^5$) Preliminary cross section.}
\label{macs}
\begin{tabular}{cccccc}
 Isotope & \multicolumn{2}{c}{Solar Abundance($^1$)} & \multicolumn{2}{c}{Hauser-Feshbach prediction [mb]} & Recommended \\
 & Anders \cite{ande89} & Lodders \cite{lodd03} &  MOST \cite{most} & NON-SMOKER \cite{rau01} & values [mb] \cite{bao00} \\
 \hline
 $^{74}$Se & 5.50$\times$10$^{-1}$ & 5.80$\times$10$^{-1}$ & 304 & 207 & \textbf{271 $\pm$ 15} \\
 $^{78}$Kr & 1.53$\times$10$^{-1}$ & 2.00$\times$10$^{-1}$ & 344 & 351 & 312 $\pm$ 26 \\
 $^{78}$Kr$\rightarrow$$^{m}$ & & & & & 92.3 $\pm$ 6.2 \\
 $^{84}$Sr & 1.32$\times$10$^{-1}$ & 1.31$\times$10$^{-1}$ & 296 ($^4$) & 393 & \textbf{300 $\pm$ 17}\\
 $^{84}$Sr$\rightarrow$$^{m}$ & & & & & \textbf{190 $\pm$ 10} \\
 $^{92}$Mo & 3.78$\times$10$^{-1}$ & 3.86$\times$10$^{-1}$ & 44 & 128 & 70 $\pm$ 10 \\
 $^{94}$Mo & 2.36$\times$10$^{-1}$ & 2.41$\times$10$^{-1}$ & 87 & 151 & 102 $\pm$ 20 \\
 $^{96}$Ru & 1.03$\times$10$^{-1}$ & 1.05$\times$10$^{-1}$ & 291 & 281 & 207 $\pm$ 8 \\
 $^{98}$Ru & 3.50$\times$10$^{-2}$ & 3.55$\times$10$^{-2}$ & 370 & 262 & 173 $\pm$ 36 ($^2$) \\
 $^{102}$Pd & 1.42$\times$10$^{-2}$ & 1.46$\times$10$^{-2}$ & 1061 & 374 & 373 $\pm$ 118 ($^2$) \\
 $^{106}$Cd & 2.01$\times$10$^{-2}$ & 1.98$\times$10$^{-2}$ & 434 & 451 & 302 $\pm$ 24 \\
 $^{108}$Cd & 1.43$\times$10$^{-2}$ & 1.41$\times$10$^{-2}$ & 260 & 373 & 202 $\pm$ 9 \\
 $^{113}$In & 7.90$\times$10$^{-3}$ & 7.80$\times$10$^{-3}$ & 413 & 1202 & 787 $\pm$ 70 \\
 $^{113}$In$\rightarrow$$^{m}$ & & & & & 480 $\pm$ 160  \\
 $^{112}$Sn & 3.72$\times$10$^{-2}$ & 3.63$\times$10$^{-2}$ & 208 & 381 & 210 $\pm$ 12 \\
 $^{114}$Sn & 2.52$\times$10$^{-2}$ & 2.46$\times$10$^{-2}$ & 106 & 270 & 134.4 $\pm$ 1.8 \\
 $^{115}$Sn & 1.29$\times$10$^{-2}$ & 1.27$\times$10$^{-2}$ & 212 & 528 & 342.4 $\pm$ 8.7 \\
 $^{120}$Te & 4.30$\times$10$^{-3}$ & 4.60$\times$10$^{-3}$ & 340 ($^4$) & 551 & \textbf{451 $\pm$ 18}($^5$) \\
 $^{120}$Te$\rightarrow$$^{m}$ & & & & & \textbf{61 $\pm$ 2}($^5$) \\
 $^{124}$Xe & 5.71$\times$10$^{-3}$ & 6.57$\times$10$^{-3}$ ($^3$) & 593 & 799 & 644 $\pm$ 83 \\
 $^{124}$Xe$\rightarrow$$^{m}$ & & & & & 131 $\pm$ 17 \\
 $^{126}$Xe & 5.09$\times$10$^{-3}$ & 5.85$\times$10$^{-3}$ ($^3$) & 472 & 534 & 359 $\pm$ 51 \\
 $^{126}$Xe$\rightarrow$$^{m}$ & & & & & 40$\pm$6 \\
 $^{130}$Ba & 4.76$\times$10$^{-3}$ & 4.60$\times$10$^{-3}$ & 561 & 730 & \textbf{694 $\pm$ 20}($^5$) \\
 $^{132}$Ba & 4.53$\times$10$^{-3}$ & 4.40$\times$10$^{-3}$ & 300 & 467 & \textbf{368 $\pm$ 25}($^5$) \\
 $^{132}$Ba$\rightarrow$$^{m}$ & & & & & \textbf{33.6 $\pm$ 1.7}($^5$) \\
 $^{136}$Ce & 2.16$\times$10$^{-3}$ & 2.17$\times$10$^{-3}$ & 227 & 495 & 328 $\pm$ 21  \\
 $^{136}$Ce$\rightarrow$$^{m}$ & & & & & 28.2 $\pm$ 1.6  \\
 $^{138}$Ce & 2.84$\times$10$^{-3}$ & 2.93$\times$10$^{-3}$ & 160 & 290 & 179 $\pm$ 5  \\
 $^{138}$La & 4.09$\times$10$^{-3}$ & 3.97$\times$10$^{-3}$ & 194 & 767 &  \\
 $^{144}$Sm & 8.00$\times$10$^{-3}$ & 7.81$\times$10$^{-3}$ & 37 & 209 & 92 $\pm$ 6   \\
 $^{156}$Dy & 2.21$\times$10$^{-3}$ & 2.16$\times$10$^{-3}$ & 2025 & 1190 & 1567 $\pm$ 145   \\
 $^{158}$Dy & 3.78$\times$10$^{-3}$ & 3.71$\times$10$^{-3}$ & 2188 & 949 & 1060 $\pm$ 400 ($^2$) \\
 $^{162}$Er & 3.51$\times$10$^{-3}$ & 3.50$\times$10$^{-3}$ & 1818 & 1042 & 1624 $\pm$ 124   \\
 $^{168}$Yb & 3.22$\times$10$^{-3}$ & 3.23$\times$10$^{-3}$ & 917 & 886 & 1160 $\pm$ 400 ($^2$)\\
 $^{174}$Hf & 2.49$\times$10$^{-3}$ & 2.75$\times$10$^{-3}$ & 709 & 786 & 956 $\pm$ 283 ($^2$) \\
 $^{180}$W  & 1.73$\times$10$^{-3}$ & 1.53$\times$10$^{-3}$ & 722 & 707 & 536 $\pm$ 60   \\
 $^{184}$Os & 1.22$\times$10$^{-3}$ & 1.33$\times$10$^{-3}$ & 697 & 789 & 657 $\pm$ 202 ($^2$) \\
 $^{190}$Pt & 1.70$\times$10$^{-3}$ & 1.85$\times$10$^{-3}$ & 659 & 760 & 677 $\pm$ 82  \\
 $^{196}$Hg & 4.80$\times$10$^{-3}$ & 6.30$\times$10$^{-3}$ & 493 & 372 & 650 $\pm$ 82 ($^2$) \\
\end{tabular}
\end{table*}

\end{document}